\documentclass[12pt]{article}

\usepackage{latexsym}

\usepackage{graphics}
\usepackage{graphicx}
\usepackage{epstopdf}
\usepackage{amssymb}
\usepackage{tabularx}
\newcommand{\be}{\begin{equation}}
\newcommand{\ee}{\end{equation}}
\newcommand{\ba}{\begin{eqnarray}}
\newcommand{\ea}{\end{eqnarray}}
\newcommand{\ban}{\begin{eqnarray*}}
\newcommand{\ean}{\end{eqnarray*}}

\begin{document}


\title{Magnetized Black Holes in an External Gravitational Field}

\author{
   Jutta Kunz$^{1}$\footnote{E-mail:\texttt{jutta.kunz@uni-oldenburg.de}},\,Petya Nedkova$^{1,2}$\footnote{E-mail: \texttt{pnedkova@phys.uni-sofia.bg}},\, Stoytcho Yazadjiev$^{2}$\footnote{E-mail:\texttt{yazad@phys.uni-sofia.bg}}\\ \\
 {\footnotesize${}^{1}$  Institut f\"{u}r Physik, Universit\"{a}t Oldenburg}\\
{\footnotesize D-26111 Oldenburg, Germany}\\
{\footnotesize${}^{2}$ Faculty of Physics, Sofia University, 5 James}\\
{\footnotesize   Bourchier Boulevard, Sofia~1164, Bulgaria }\\}
\date{}
\maketitle

\begin{abstract}
We obtain a family of exact solutions describing magnetized black holes in an external gravitational field. Locally the solutions can be interpreted as representing the near-horizon region of a black hole, which interacts with a surrounding matter distribution producing a strong magnetic field. Thus, the solutions reflect the influence of both a gravitational and an electromagnetic external potential in the strong field regime. The static members in the family are generalizations of the Schwarzschild solution in the described environment, while the rotating ones generalize the magnetized Reissner-Nordstr\"{o}m solution when the influence of an external gravitational source is also taken into account. Technically, the solutions are obtained by means of a Harrison transformation.  We examine the thermodynamical properties of the solutions, and compare them with the corresponding isolated black holes, and with the particular cases when the interaction with only one of the external potentials is taken into account. For the static black holes the influence of the external gravitational and magnetic fields is factorized in a sense, both affecting different properties, and leaving the rest intact. For the rotating solutions the external gravitational and magnetic fields are coupled through the conditions for avoiding conical singularities. The Meissner effect is observed for extremal rotating solutions only in the zero-charge limit, similar to the magnetized Reissner-Nordstr\"{o}m black hole.
\end{abstract}

\section{Introduction}

The stationary electrovacuum black hole solutions with flat asymptotics were studied thoroughly within classical general relativity, and many of their properties were elucidated. Although of certain fundamental importance, these solutions describe very idealized physical situations. They represent isolated objects in equilibrium, which do not interact with their environment, and any physical fields are perceived as propagating in their background. In an attempt to provide descriptions of some astrophysically more relevant scenarios, two families of asymptotically non-flat black hole solutions were considered. The family of distorted black holes \cite{Geroch} includes a quasi-stationary interaction with an external gravitational source. They are interpreted as local solutions describing the near-horizon region of a quasi-equilibrium black hole and matter configuration, such as a black hole and an accretion disk, or a black hole in a binary system in the very initial stages of the inspiral. The other family of solutions, the magnetized black holes \cite{Aliev}, involve interaction with a stationary magnetic field aligned with the symmetry axis. The magnetic field is interpreted as produced by some external source, such as a surrounding accretion disk, and not as an intrinsic characteristic of the black hole. Both families of solutions describe the strong field regime of gravitational and electromagnetic interactions, taking into account the non-linear back-reaction of the black hole geometry  on the external potential.

The idea of interpreting the distorted black holes as interacting with an external gravitational field was introduced in the work of Geroch and Hartle \cite{Geroch}. They  constructed the most general static vacuum black hole solution with a regular event horizon in classical general relativity. The solution is not asymptotically flat if considered as a global solution,  and the deviation from asymptotic flatness is attributed to the influence of an external gravitational potential. The Geroch-Hartle solution was generalized to charged and rotating black holes by means of solution generation transformations \cite{Tomimatsu}-\cite{Breton:1998}, and extended to higher dimensions \cite{Abdolrahimi:2010}-\cite{Nedkova:2014}. Recent work demonstrated that distorted black holes can possess properties, which are qualitatively different from their isolated counterparts. For example, they can violate the Kerr bound of the angular momentum/mass ratio, without leading to the formation of naked singularities \cite{Abdolrahimi:2015}.

The magnetized black holes were introduced by considering a non-linear superposition of the Schwarzschild black hole and the Melvin's magnetic universe \cite{Ernst:1976a}. The resulting solution describes a black hole situated in a large-scale self-gravitating magnetic field, which is approximately uniform at large distances. The solution does not possess a magnetic charge, but is characterized by a non-trivial magnetic flux through the horizon, which is considered to be  generated by an external source. Therefore, such configurations are interpreted as black holes interacting with an external magnetic field. The Schwarzschild-Melvin solution constitutes the unique static axisymmetric black hole solution of the Einstein-Maxwell equations, which possesses the Melvin asymptotics \cite{Hiscock}. Rotating and charged magnetized black holes, constructed as its generalizations, are characterized with much more complicated asymptotic behavior \cite{Ernst:1976b}-\cite{Diaz}. The interaction of the gravitational and magnetic fields causes a global rotation of the spacetime, which can be avoided only in special cases. This leads to non-uniqueness in the choice of the asymptotically timelike Killing vector, resulting in possible qualitatively different ergoregion configurations \cite{Gibbons:2013}. The definition of global charges for stationary magnetized black holes is also a non-trivial problem, and different constructions proposed recently lead to non-equivalent results \cite{Gibbons:2013b}, \cite{Compere}.

The magnetized black holes received much attention in an astrophysical context. The magnetic field produced by the accretion disk is considered to be a crucial ingredient in astrophysical processes, such as the formation of relativistic jets from galactic nuclei and microquasars. One of the most relevant models for jet formation, the Blandford-Znajek mechanism, suggest that the magnetic lines threading the horizon are dragged by the rotating black hole, pushing plasma along the black hole rotation axis in the form of two opposite jets. The power radiated from the black hole is proportional to the spin parameter, and achieves its highest values for extremal rotation. In this regime, however, a counteracting mechanism is activated. Magnetic field lines are expelled from the extremal horizons, similar to the Meissner effect for superconductors, quenching the formation of relativistic jets. It is still an open question whether jets can be produced by black holes with near-extremal rotation.

The Meissner effect is considered a fundamental property of the extremal horizons, connected with the fact that the length of the black hole throat gets infinite in the extremal limit, and the modes on the both sides of the horizon become disentangled \cite{Penna:2014a}. In support of the latter interpretation, it was derived from the properties of the Hartle-Hawking vacuum at the low temperature limit \cite{Penna:2014b}. The Meissner effect is observed for extremal black holes in various gravitational theories \cite{Emparan}. However, it can be suppressed in certain physical situations, as for example when the external magnetic field is not aligned with the symmetry axis \cite{Bicak:1985}, or the black hole possesses intrinsic electric charge \cite{Emparan}. In such cases the magnetic flux through the extremal horizon does not vanish.

The purpose of this paper is to construct exact solutions which reflect the influence of both external gravitational and magnetic potentials. The solutions model to some extent relevant astrophysical situations, such as a black hole surrounded by a massive accretion disk, which produces a strong magnetic field. They are also of fundamental importance since they provide intuition how the described gravi-magnetic interaction modifies the black hole properties.

The magnetized black holes are constructed technically by performing a Harrison transformation on an appropriate seed solution \cite{Harrison}. For our purposes we choose seeds, which represent distorted black holes. Thus, the resulting solutions are interpreted as local solutions describing the near-horizon region of black holes, which interact with an external gravitational potential and a large-scale magnetic field, not produced by the central object. We realize the described idea by constructing a static family of solutions, which generalize the Schwarzschild black hole. The solution reduces to the magnetized Schwarzschild black hole in the limit when the external gravitational potential vanishes, and to the Geroch-Hartle distorted black hole when the external magnetic field is switched off. In the class of magnetized black holes obtained by a Harrison transformation, this solution represents the most general static black holes with a regular horizon. As an example for a stationary solution, we construct a generalization of the Reissner-Nordstr\"{o}m black hole. The solution is obtained by performing a Harrison transformation on the distorted Reissner-Nordstr\"{o}m black hole. The Harrison transformation applied on a charged seed generates rotation, thus the constructed solution is stationary, rather than static.

The paper is organized as follows. In section 2 we review the Schwarzschild and the Reissner-Nordstr\"{o}m black holes in an external gravitation field.  We represent the solutions in prolate spheroidal coordinates for convenience for further calculations. In section 3 we describe the Harrison transformation used for the construction of magnetized black holes. We consider in detail the case when the Harrison transformation is performed on a static electrovacuum seed.  In section 4 and 5 we obtain the magnetized Schwarzschild and Reissner-Nordstr\"{o}m black holes in an external gravitational field and examine their thermodynamical properties. In the construction of the magnetized distorted Reissner-Nordstr\"{o}m black hole we derive general expressions for the metric functions and electromagnetic potentials, resulting from the application of the Harrison transformation on an arbitrary electrovacuum Weyl seed. We discover certain differential relations valid for the electrovacuum Weyl class, which facilitate considerably the integration procedure. Section 6 contains conclusions.

\section{Static black holes in an external gravitational field}

Currently, two families of static solutions describing black holes in an external gravitational fields are obtained in the classical general relativity, which belong to the class of (electro-) vacuum Weyl solutions.  A  vacuum solution, which generalizes the Schwarzschild black hole in the presence of a gravitational interaction, was constructed in \cite{Geroch}, and an electrovacuum one corresponding to the Reissner-Nordstr\"{o}m solution was obtained in \cite{Fairhurst}. While the former represents the most general distorted vacuum black hole with a regular event horizon, the latter is a particular solution describing electrovacuum distorted black holes, which possess a similar structure as the Reissner-Nordstr\"{o}m solution.

\subsection{Schwarzschild black hole in an external gravitational field}

The Schwarzschild black hole in an external gravitational field is a static axisymmetric solution to the Einstein equations in vacuum with non-flat asymptotics. Its metric is given in the prolate spheroidal coordinates $(x,y)$ in the form

\begin{eqnarray}\label{dist_Sch}
ds^2 &=& -e^{2\psi_0} dt^2 +  \sigma^2e^{-2\psi_0} \left[e^{2\gamma_0}  \left(\frac{dx^2}{x^2-1}+\frac{dy^2}{1-y^2}\right) + (x^2-1)(1-y^2)d\phi^2\right], \nonumber \\[2mm]
\psi_0 &=& \psi_{S} + \widetilde{\psi_0}, \quad~~~
\psi_S = \frac{1}{2}\ln{\left(\frac{x-1}{x+1}\right)}, \nonumber\\[2mm]
{\widetilde\psi_0} &=& \sum^{\infty}_{n=0}a_nR^n P_n\left(\frac{xy}{R}\right), \quad ~~~ R = \sqrt{x^2 +  y^2 - 1},
\end{eqnarray}
where $x\geq 1$, and $-1\leq y\leq 1$ in the domain of outer communication. The physical infinity is located at $x\rightarrow\infty$. The function $\psi_0$ is a harmonic function in an auxiliary non-physical flat space. It represents a superposition of the potential $\psi_S$ corresponding to the isolated Schwarzschild black hole, and  the potential $\widetilde\psi_0$, which characterizes the external gravitational source. The potential $\widetilde\psi_0$ is expressed in terms of the Legendre polynomials $P_n$, and is determined by a discrete set of real constants $a_n, \, n \in\mathcal{N}$. The metric function $\gamma_0$ is a solution to the linear system

\begin{eqnarray}\label{gamma_0}
\partial_x\gamma_0 &=& \frac{1-y^2}{x^2-y^2}\left[x(x^2-1)(\partial_x\psi_0)^2 - x(1-y^2)(\partial_y\psi_0)^2 - 2y(x^2-1)\partial_x\psi_0\,\partial_y\psi_0\right], \nonumber \\[2mm]
\partial_y\gamma_0 &=& \frac{x^2-1}{x^2-y^2}\left[y(x^2-1)(\partial_x\psi_0)^2 - y(1-y^2)(\partial_y\psi_0)^2 + 2x(1-y^2)\partial_x\psi_0\,\partial_y\psi_0\right].
\end{eqnarray}

It can also be represented as a contribution from the isolated Schwarzschild solution $\gamma_S$ and a term $\widetilde \gamma_0$ introduced by the interaction with the external gravitational field

\begin{eqnarray}\label{gamma_Sch}
\gamma_0 &=& \gamma_S + \widetilde\gamma_0, \quad \gamma_S = \frac{1}{2}\ln(x^2-1), \\[2mm]
\widetilde\gamma_0 &=& \sum^{\infty}_{n,k=1}\frac{nk}{n+k}a_na_k R^{n+k}\left(P_n P_k - P_{n-1}P_{k-1}\right) \nonumber \\
&+& \sum^{\infty}_{n=1}a_n\sum^{n-1}_{k=0}\left[(-1)^{n-k+1}(x+y) -x+y\right]R^k P_k\left(\frac{xy}{R}\right). \nonumber
\end{eqnarray}

The prolate spheroidal coordinates are connected to the conventional Schwarzschild coordinates $(r, \theta)$ by the relations

\begin{eqnarray}\label{coord_Schw}
x = \frac{r}{\sigma}-1, \quad~~~ y = \cos\theta.
\end{eqnarray}

The solution contains a Killing horizon located at $x=1$, and the symmetry axis consists of two disconnected components  at $y=1$ and $y=-1$. It is characterized by the real parameter $\sigma$, which is equal to the Komar mass on the black hole horizon, and the discrete set of constants $a_n, \, n\in \mathcal{N}$, which determine the external gravitational field. The isolated Schwarzschild black hole is recovered in the limit when all the parameters $a_n$ vanish. For balanced solutions the external matter is restricted by the condition for absence of conical singularities on the axis. It reduces to the following constraint on the solution parameters

\begin{eqnarray}\label{balance_Sch}
\sum^{\infty}_{n=0}a_{2n+1} =0.
\end{eqnarray}

The interaction with the external potential leads to deformation of  the horizon geometry with respect to the spherical one. The metric on the horizon cross-section is given by

\begin{eqnarray}\label{hor_metricDS}
ds_H^2 &=&  4\sigma^2e^{\widetilde\gamma_0(y)-2\widetilde\psi_0(y)}\left[e^{-\widetilde\gamma_0(y)}(1-y^2)d\phi^2 + e^{\widetilde\gamma_0(y)}\frac{dy^2}{1-y^2}\right]  \\[2mm]
 &=& 4\sigma^2 e^{\widetilde\gamma_0(\theta)-2\widetilde\psi_0(\theta)}\left[e^{-\widetilde\gamma_0(\theta)}\sin^2\theta d\phi^2 + e^{\widetilde\gamma_0(\theta)}d\theta^2\right], \quad~~~ y = \cos\theta \nonumber
\end{eqnarray}
As a result of the field equations ($\ref{gamma_0}$), the combination of metric functions $\widetilde\psi_0 - \frac{1}{2}\widetilde\gamma_0$ reduces to a constant on the horizon, which is equal to the value of the metric function $\widetilde\psi_0$ at the intersection of the horizon with the rotational axis $x=1, \, y=\pm 1$. Hence, the horizon geometry is determined by a single function $\widetilde\psi_0(y)$ (or alternatively $\widetilde\gamma_0(y)$), controlling the deviation from the geometrical sphere, which can be interpreted as a shape function \cite{Hajicek:1973}. It restricts to the expression

\begin{eqnarray}
\widetilde\psi_0|_H = \sum^{\infty}_{n=0}a_n y^n, \quad ~~~ -1\leq y \leq 1,
\end{eqnarray}
on the horizon, and for balanced solutions takes equal values on the horizon intersection with the two axis components $\widetilde\psi_0(x=1, y=\pm 1) = \sum^{\infty}_{n=0}a_{2n}$. The horizon area takes the form

\begin{eqnarray}
A_H &=& 16\pi\sigma^2e^{-2\widetilde\psi_0}|_{x=1, y=\pm 1} = 4\pi R_H^2, \\ \nonumber
R_H &=& 2\sigma\exp{\left[-\sum^{\infty}_{n=0}a_{2n}\right]},
\end{eqnarray}
The scalar quantity $R_H$, appearing also as a scale factor in ($\ref{hor_metricDS}$), is interpreted as an effective horizon radius in analogy with the spherical case.

\subsection{Reissner-Nordstr\"{o}m black hole in an external gravitational field}

The Reissner-Nordstr\"{o}m black hole in an external gravitational field is a static electrovacuum solution with non-flat asymptotics. It contains the Reissner-Nordstr\"{o}m black hole as a limiting case, when asymptotic flatness is achieved, and the distorted Schwarzschild solution in the limit, when its charge vanishes. The solution belongs to the electrovacuum Weyl class of solutions \cite{Weyl}. It is characterized by two potentials, a gravitational and an electromagnetic one, which are assumed to be related by a  functional dependence. Due to this relation the electrovacuum field equations acquire the same form as for the vacuum Weyl class. They reduce to a Laplace equation for a certain generalized potential, and a decoupled linear system corresponding to ($\ref{gamma_0}$). This observation is used to construct a map between vacuum and elecrovacuum Weyl solutions and to generate charged solutions by applying an algebraic transformation on a vacuum seed \cite{Weyl},\cite{Gatreau:1972}.

Describing electrovacuum solutions, we consider the general form of the metric for the Weyl class in the prolate spheroidal coordinates

\begin{eqnarray}\label{dist_RN}
ds^2 = -e^{2\psi} dt^2 +  \sigma^2e^{-2\psi}\left[e^{2\gamma} \left(\frac{dx^2}{x^2-1}+\frac{dy^2}{1-y^2}\right) + (x^2-1)(1-y^2)d\phi^2\right],
\end{eqnarray}
and denote the electromagnetic field by $F = d\chi\wedge dt$. If we assume a functional relation between the gravitational and electromagnetic potentials $\psi = \psi(\chi)$, it follows from the field equations that it should possesses the explicit form

\begin{eqnarray}\label{chi_Weyl}
e^{2\psi} = 1 - 2C\chi + \chi^2,
\end{eqnarray}
where $C$ is an arbitrary constant. The parameter $C$ is interpreted physically as the ratio between the charge and the mass of the gravitational source. We consider a vacuum Weyl solution characterized by a gravitational potential $\psi_0$. Then, using the map which we discussed, it determines the gravitational potential of an electrovacuum Weyl solution by the relation

\begin{eqnarray}\label{psi_Weyl}
e^{-\psi} = \frac{1}{2}\left[\left(1+ \frac{C}{\sqrt{C^2-1}}\right)e^{-\psi_0} + \left(1-  \frac{C}{\sqrt{C^2-1}}\right)e^{\psi_0}\right].
\end{eqnarray}

The map between the two solutions preserves the metric function $\gamma$, i.e. it will possess the same form for the constructed charged solution as for the vacuum seed.

The Reissner-Nordstr\"{o}m black hole can be obtained by applying the described transformation on the Schwarzschild solution. A charged black hole in an external gravitational field can be constructed in the same way by applying the transformation on the distorted Schwarzschild black hole ($\ref{dist_Sch}$) instead. The solution was obtained originally in \cite{Fairhurst}, however for our purposes we will present it in a different form. The metric is given by the general expression in the prolate spheroidal coordinates ($\ref{dist_RN}$), and the gravitational potential is expressed by
\begin{eqnarray}\label{psi}
e^{\psi} = \frac{2\sigma e^{\psi_0}}{\sigma + M + (\sigma-M)e^{2\psi_0}},
\end{eqnarray}
where $\psi_0$ is the potential for the distorted Schwarzschild solution ($\ref{dist_Sch}$). It contains an asymptotically non-flat part,  which describes the interaction with an external gravitational source. The remaining metric  function satisfies $\gamma = \gamma_0$, where $\gamma_0$ is given by ($\ref{gamma_0}$), and the electromagnetic field can be represented in the form

\begin{eqnarray}\label{chi}
F &=& d\chi\wedge dt, \nonumber \\[2mm]
\chi &=& \frac{Q(1-e^{2\psi_0})}{\sigma + M + (\sigma - M)e^{2\psi_0}}.
\end{eqnarray}

The solution contains a Killing horizon located at $x=1$, and the symmetry axis corresponds to $y=1$ and $y=-1$. It is characterized by the real parameters $M$ and $Q$, which are related to its mass and charge. Their ratio is equal to the parameter $C=M/Q$ introduced by the Weyl transformation. The real constant $\sigma$ is the mass parameter of the vacuum seed solution, which satisfies the relation $\sigma^2 = M^2-Q^2 = Q^2(C^2-1)$. It is also equal to the half-length of the horizon interval in the factor space of the solution manifold with respect to the isometry group. Considering the relations between the solution parameters, we can see that the expressions for the gravitational and electromagnetic potentials are equivalent to the general expressions ($\ref{chi_Weyl}$)-($\ref{psi_Weyl}$). The interaction with the external gravitational field is encoded in the asymptotically non-flat part of the seed potential $\psi_0$. As in the vacuum case, it is determined by a set of real constants $a_n, \, n\in \mathcal{N}$. In the limit when all the constants $a_n$ vanish, and $\psi_0$ reduces to the potential of the isolated Schwarzschild solution, we obtain the isolated Reissner-Nordstr\"{o}m solution. The Schwarzschild black hole in an external gravitational field, which we descried in the previous section, is recovered when we set $Q=0$. The solution is in equilibrium when the external matter parameters satisfy the same balance condition ($\ref{balance_Sch}$) as for the distorted Schwarzschild solution.

The prolate spheroidal coordinates are connected to the conventional Reissner-Nordstr\"{o}m coordinates $(r, \theta)$ by the relation

\begin{eqnarray}
x = \frac{r-M}{\sigma}, \quad~~~ y = \cos\theta.
\end{eqnarray}

The external gravitational field deforms the horizon from the spherical geometry in a similar way as for the distorted Schwarzschild solution. The horizon geometry is determined by the metric

\begin{eqnarray}
ds_H^2 &=& (\sigma + M)^2 e^{\widetilde\gamma_0(y)-2\widetilde\psi_0(y)}\left[e^{-\widetilde\gamma_0(y)}(1-y^2)d\phi^2 + e^{\widetilde\gamma_0(y)}\frac{dy^2}{1-y^2}\right]  \\[2mm]
 &=& (\sigma + M)^2 e^{\widetilde\gamma_0(\theta)-2\widetilde\psi_0(\theta)}\left[e^{-\widetilde\gamma_0(\theta)}\sin^2\theta d\phi^2 + e^{\widetilde\gamma_0(\theta)}d\theta^2\right], \quad~~~ y = \cos\theta, \nonumber
\end{eqnarray}
where the metric functions $\widetilde\psi_0$ and $\widetilde\gamma_0$ characterize the distorted Schwarzschild solution. The horizon area is given by the expression
\begin{eqnarray}
A_H &=& 4\pi(\sigma + M)^2 e^{-2\widetilde\psi_0}|_{x=1, y=\pm 1} = 4\pi R_H^2, \\ \nonumber
R_H &=& (\sigma + M)\exp{\left[-\sum^{\infty}_{n=0}a_{2n}\right]},
\end{eqnarray}
where we have introduced an effective horizon radius $R_H$. Hence, the horizon is deformed by the same shape function as for the distorted Schwarzschild solution, and the horizon radius deviates in the same way from its value in the isolated case.

\section{Construction of magnetized black holes}

We consider stationary axisymmetric solutions of the Einstein-Maxwell equations. The metric is expressed in the prolate spheroidal coordinates in terms of the gravitational potential with respect to the spacelike Killing field as

\begin{eqnarray}\label{metric_u}
ds^2 = e^{2u}(d\phi + \omega dt)^2 + \sigma^2e^{-2u}\left[e^{2\gamma} \left(\frac{dx^2}{x^2-1}+\frac{dy^2}{1-y^2}\right) - (x^2-1)(1-y^2) dt^2\right],
\end{eqnarray}
and the electromagnetic field is in the form
\begin{eqnarray}
F = dA_t\wedge dt + dA_\phi\wedge d\phi,
\end{eqnarray}
where all the functions depend on the prolate spheroidal coordinates $(x,y)$. The problem can be equivalently described by introducing two complex potentials, a gravitational one ${\cal E}$ and an electromagnetic potential $\Phi$, defined as \cite{Ernst:1968}

\begin{eqnarray}
{\cal E} &=& - e^{2u} - |\Phi|^2 + if, \nonumber\\
\Phi &=& A_\phi + iB_\phi,
\end{eqnarray}
We denote by $f$ the twist potential, and $B_{\phi}$ is the dual electromagnetic potential. It is determined by means of the Maxwell 2-form $F$ as $dB_{\phi} = i_\eta \star F$, where $\eta$ is the Killing vector generating the azimuthal symmetry $\eta=\frac{\partial}{\partial\phi}$. Using the complex potentials the stationary and axisymmetric Einstein-Maxwell equations can be reduced to two nonlinear Ernst equations for ${\cal E}$ and $\Phi$ \cite{Ernst:1968}. The Ernst equations are invariant under the group $SU(2,1)$ and solutions can be constructed by its action on the potential space. The group transformation generates a new pair of complex potentials, which determine a new solution to the Einstein-Maxwell equations. A particular 1-parameter transformation obtained by Harrison \cite{Harrison} was observed to lead to solutions containing a large-scale magnetic field \cite{Ernst:1976a}-\cite{Ernst:1976b}. Its parameter, which we will denote by $B$, is associated with the strength of the magnetic field at infinity. These solutions are interpreted as located in an external magnetic field. If we consider an already known seed solution to the Einstein-Maxwell equations with potentials ${\cal E}_0$ and $\Phi_0$, the Harrison transformation generates  new potentials by the relations \cite{Ernst:1976a}

\begin{eqnarray}
{\cal E} &=& \Lambda^{-1}{\cal E}_0, \nonumber \\
\Phi &=& \Lambda^{-1}(\Phi_0 - \frac{1}{2}B{\cal E}_0),
\end{eqnarray}

\noindent
by means of a complex function
\begin{eqnarray}
\Lambda = 1 + B\Phi_0 - \frac{1}{4}B^2{\cal E}_0.
\end{eqnarray}

The metric functions and the electromagnetic field are further extracted from these expressions. The gravitational potential $u$ is determined from the real part of the Ernst potential ${\cal E}$ in the form
\begin{eqnarray}
e^{2u}= |\Lambda|^{-2}e^{2u_0},
\end{eqnarray}
where $u_0$ is the corresponding potential for the seed solution. The metric function $\omega$ is obtained as a solution to the equations

\begin{eqnarray}\label{omega_Mgn}
\frac{\partial\omega}{\partial x} &=& \frac{2\sigma(1-y^2)}{e^{2u_0}}\left[{\mathrm{Im}}\Lambda\,\partial_y(\mathrm{Re}\Lambda) - \mathrm{Re}\Lambda\,\partial_y(\mathrm{Im}\Lambda)\right], \nonumber \\[2mm]
\frac{\partial\omega}{\partial y} &=& -\frac{2\sigma(x^2-1)}{e^{2u_0}}\left[\mathrm{Im}\Lambda\,\partial_x(\mathrm{Re}\Lambda) - \mathrm{Re}\Lambda\,\partial_x(\mathrm{Im}\Lambda)\right].
\end{eqnarray}

The Harrison transformation leaves the metric function $\gamma$ invariant, i.e. it is satisfied that $\gamma = \gamma_0$. As a result we obtain the metric of the new solution in the form

\begin{eqnarray}\label{metric_magn}
ds^2 &=& |\Lambda|^{-2}e^{2u_0}(d\phi + \omega dt)^2 \nonumber \\[2mm]
&+& |\Lambda|^2 \sigma^2e^{-2u_0}\left[e^{2\gamma_0} \left(\frac{dx^2}{x^2-1}+\frac{dy^2}{1-y^2}\right) - (x^2-1)(1-y^2) dt^2\right].
\end{eqnarray}

In the following we will consider only static electrically charged seed solutions, i.e. possessing an electromagnetic field in the form $F_0 = dA_t^0\wedge dt$. The electromagnetic potential $A_\phi$ and the dual potential $B_\phi$ of the magnetized solution can be extracted directly from the real and imaginary parts of the complex potential $\Phi$. They are given by the relations

\begin{eqnarray}\label{potential_magn}
A_\phi &=& \frac{B}{2|\Lambda|^2} \left[-{\cal E}_0 + \frac{1}{4}B^2{\cal E}^2_0 + 2(B^0_\phi)^2\right], \nonumber \\[2mm]
B_\phi &=& \frac{B^0_\phi}{|\Lambda|^2}\left[ 1 + \frac{1}{4}B^2{\cal E}_0\right],
\end{eqnarray}
where $B_\phi^0$ is the dual electromagnetic potential of the seed solution, and ${\cal E}_0$ is its Ernst potential. In order to obtain the potential $A_t$, it is convenient to consider the co-rotating potential $\Sigma$, defined as $\Sigma = A_t + \omega A_\phi$. It is related to the dual potential $B_\phi$ by the equations
\begin{eqnarray}
\frac{\partial \Sigma}{\partial x} &=& \frac{\sigma(y^2-1)}{e^{2u}}\frac{\partial B_\phi}{\partial y} + A_\phi\,\frac{\partial\omega}{\partial x}, \nonumber \\[2mm]
\frac{\partial \Sigma}{\partial y} &=& \frac{\sigma(x^2-1)}{e^{2u}}\frac{\partial B_\phi}{\partial x} + A_\phi\,\frac{\partial\omega}{\partial y}.
\end{eqnarray}
These equations can be reduced to the relation
\begin{eqnarray}\label{Sigma_gen}
d\Sigma = \frac{3}{2B}d\omega + 2dA^0_t,
\end{eqnarray}
which allows to obtain the explicit form of the potential $A_t$ generated by the Harrison transformation
\begin{eqnarray}
A_t &=& 2A^0_t -\omega A_\phi + \frac{3}{2B}\omega + const.\, ,
\end{eqnarray}
where $A^0_t$ is the corresponding potential for the seed solution.
We should note that the Harrison transformation generates static solutions only from vacuum seeds. If we consider an electrovacuum seed solution, it introduces rotation  due to the interaction of its electromagnetic potential with the external electromagnetic field.

\section{Magnetized Schwarzschild black hole in an external gravitational field}

Static magnetized black holes are obtained in a particularly simple way by means of the Harrison transformation. In this case the transformation is applied on a static vacuum seed, and it reduces purely to algebraic operations. The solution is characterized only by the Ernst potential, which is a real function, as well as the metric function $\Lambda$ following from it. In this section we construct a magnetized Schwarzschild black hole in an external gravitational field. This is achieved by applying the Harrison transformation on the distorted Schwarzschild solution described in section 2. The seed metric ($\ref{dist_Sch}$) is represented in an equivalent form by means of the gravitational potential with respect to the spacelike Killing field
\begin{eqnarray}\label{metric_u}
ds^2 = e^{2u_0}d\phi^2 + \sigma^2e^{-2u_0}\left[e^{2\gamma^{'}_0} \left(\frac{dx^2}{x^2-1}+\frac{dy^2}{1-y^2}\right) - (x^2-1)(1-y^2) dt^2\right],
\end{eqnarray}
where the potential $u_0$ and the metric function $\gamma^{'}_0$ are related to the original quantities $\psi_0$ and $\gamma_0$ as

\begin{eqnarray}\label{potential_u_0}
e^{2u_0}&=&\sigma^2(x^2-1)(1-y^2)e^{-2\psi_0}, \nonumber \\
e^{2\gamma^{'}_0} &=& \sigma^2(x^2-1)(1-y^2)e^{2\gamma_0-4\psi_0}.
\end{eqnarray}
Then, the Ernst potential is given by

\begin{eqnarray}
{\cal E}_0 &=& -e^{2u_0}, \nonumber \\
\end{eqnarray}
and the magnetized solution is constructed straightforwardly
\begin{eqnarray}
ds^2 &=& \Lambda^{-2}e^{2u_0}d\phi^2 + \nonumber \\[2mm]
&+& \Lambda^2 \sigma^2e^{-2u_0}\left[e^{2\gamma^{'}_0} \left(\frac{dx^2}{x^2-1}+\frac{dy^2}{1-y^2}\right) - (x^2-1)(1-y^2) dt^2\right], \nonumber \\[2mm]
\Lambda &=& 1-\frac{1}{4}B^2{\cal E}_0 = 1 + \frac{1}{4}B^2\sigma^2(x^2-1)(1-y^2)e^{-2\psi_0}.
\end{eqnarray}

\noindent
according to ($\ref{metric_magn}$) and ($\ref{potential_u_0}$), where the functions $\psi_0$ and $\gamma_0$ characterize the seed solution
\begin{eqnarray}
\psi_0 &=& \frac{1}{2}\ln{\left(\frac{x-1}{x+1}\right)} + \sum^{\infty}_{n=0}a_nR^n P_n\left(\frac{xy}{R}\right), \nonumber\\[2mm]
\gamma_0 &=&  \frac{1}{2}\ln(x^2-1)+ \sum^{\infty}_{n,k=1}\frac{nk}{n+k}a_na_k R^{n+k}\left(P_n P_k - P_{n-1}P_{k-1}\right) \nonumber \\
&+& \sum^{\infty}_{n=1}a_n\sum^{n-1}_{k=0}\left[(-1)^{n-k+1}(x+y) -x+y\right]R^k P_k\left(\frac{xy}{R}\right). \nonumber
\end{eqnarray}

The electromagnetic field is given by
\begin{eqnarray}
F &=& dA_\phi\wedge d\phi,  \nonumber \\
A_\phi &=& \frac{B}{2\Lambda}\sigma^2(x^2-1)(1-y^2)e^{-2\psi_0},  \nonumber
\end{eqnarray}
following the general relations ($\ref{potential_magn}$).

\subsection{Properties}

The solution possesses a Killing horizon located at $x=1$, and the symmetry axis corresponds to $y = \pm 1$. It contains the magnetized Schwarzschild black hole \cite{Ernst:1976a} in the limit $a_n =0, \, n\in {\cal N}$, which acquires its conventional form by means of the coordinate transformation ($\ref{coord_Schw}$). The solution is free of conical singularities provided that the norm $K = g(\eta,\eta)$ of the spacelike Killing vector $\eta=\partial/\partial\phi$ satisfies the condition

\begin{eqnarray}\label{conicalG}
\frac{1}{4K}g^{\mu\nu}\partial_\mu K\partial_\nu K \longrightarrow 1,
\end{eqnarray}
in the vicinity of the rotational axis. Imposing this condition we assume that the angular coordinate $\phi$ possesses the standard periodicity $\Delta\phi = 2\pi$. Explicit calculation shows that it coincides with the regularity condition for the seed solution

\begin{eqnarray}
\sum^{\infty}_{n=0}a_{2n+1} =0.
\end{eqnarray}

\noindent
The surface gravity of the horizon is defined as
\begin{eqnarray}
\kappa^2_H = -\frac{1}{4\lambda} g^{\mu\nu}\partial_{\mu}\lambda
\partial_{\nu}\lambda ,
\end{eqnarray}

\noindent
where $\lambda=g(\chi,\chi)$ is the norm of the Killing field $\chi = \partial/\partial t$.  We obtain again the same expression as for the distorted Schwarzschild black hole

\begin{eqnarray}
\kappa_H =\frac{1}{4\sigma} \exp\left[2\sum_{n=1}^\infty a_{2n}\right].
\end{eqnarray}

The black hole is further characterized by a local mass $M_H$ defined by the Komar integral evaluated on the horizon

\begin{eqnarray}\label{Komar}
M_{H} &=&  - {1\over 8\pi} \int_{H} \star d\xi = \sigma ,
\end{eqnarray}
where $\xi = \partial/\partial t$. The Komar mass on the horizon proves to be unaffected either by the external gravitational potential, or by the external magnetic field retaining the same value as for the isolated Schwarzschild black hole.

We examine the deformation in the horizon geometry resulting from the influence of the external magnetic field. The restriction of the metric on the horizon cross-section is given by
\begin{eqnarray}\label{hor_metricMS}
ds_H^2 &=&  4\sigma^2e^{\widetilde\gamma_0(y)-2\widetilde\psi_0(y)}\left[\Lambda^{-2}e^{-\widetilde\gamma_0(y)}(1-y^2)d\phi^2 + \Lambda^2 e^{\widetilde\gamma_0(y)}\frac{dy^2}{1-y^2}\right]  \\[2mm]
 &=& 4\sigma^2 e^{\widetilde\gamma_0(\theta)-2\widetilde\psi_0(\theta)}\left[\Lambda^{-2}e^{-\widetilde\gamma_0(\theta)}\sin^2\theta d\phi^2 + \Lambda^2e^{\widetilde\gamma_0(\theta)}d\theta^2\right], \quad~~~ y = \cos\theta \nonumber \\[2mm]
\end{eqnarray}
where the metric function $\Lambda$ reduces to

\begin{eqnarray}
\Lambda = 1 + \sigma^2 (1-y^2)e^{-2\widetilde\psi_0(y)} = 1 + \sigma^2\sin^2\theta e^{-2\widetilde\psi_0(\theta)}.
\end{eqnarray}
The combination of metric functions $\widetilde\psi_0 - \frac{1}{2}\widetilde\gamma_0$ is a constant on the horizon

\begin{eqnarray}
\widetilde\psi_0 - \frac{1}{2}\widetilde\gamma_0|_H = \sum^{\infty}_{n=0}a_{2n},
\end{eqnarray}
as discussed in section 2, and the scale factor in ($\ref{hor_metricMS}$) is interpreted as an effective horizon radius $R_H$. We see that the horizon radius is not influenced by the magnetic field, and coincides with that of the vacuum seed solution ($\ref{dist_Sch}$)

\begin{eqnarray}
R_H &=& 2\sigma\exp{\left[-\sum^{\infty}_{n=0}a_{2n}\right]}.
\end{eqnarray}
As a result, the horizon area $A_H = 4\pi R_H^2$  is also preserved. The influence of the magnetic field is encoded in the shape function, controlling the deformation of the horizon from the geometrical sphere, which is modified by the factor $\Lambda^2$.

The constructed solution does not possess an electric or magnetic charge, as can be proven by direct calculation. Its electromagnetic properties are characterized by the magnetic flux through the horizon upper/lower hemisphere

\begin{eqnarray}
F_B = \frac{1}{4\pi}\int_{H^+} F = \frac{\Delta\phi}{4\pi} \left[ A_\phi(x=1, y=0) - A_\phi(x=1, y=1)\right].
\end{eqnarray}

As a result of the calculation we obtain the same expression as for the magnetized Schwarzschild black hole \cite{Bicak:1985}

\begin{eqnarray}
F_B = \frac{\sigma^2 B}{1 + B^2\sigma^2},
\end{eqnarray}
showing that the external gravitational field has no influence on the electromagnetic properties of the solution.

In summary we can conclude that the influence of the external gravitational field and that of the external magnetic field are factorized in a sense. Some properties are modified by the former, but not by the latter, and the opposite situation is also observed. Certain characteristic like the local mass of the black hole are not influenced by either of the external potentials. We overview the examined properties in table $\ref{static}$, where we compare their values for the isolated Schwarzschild black hole (Isol. Schw.), its generalizations in an external gravitational filed \cite{Geroch} (Dist. Schw.), and in an external magnetic field \cite{Ernst:1976a} (Magn. Schw.), and for the constructed solution in this section (Magn. Dist. Schw.), which reflects the influence of both external potentials.

\begin{table}[h]
\begin{center}
\begin{tabular}{|c|c|c|c|c| }
\hline
 &\rule{0pt}{2ex} Isol. Schw. &  Dist. Schw.& Magn. Schw.&  Magn. Dist. Schw.\\[1mm]
\hline
\rule{0pt}{3ex}   Surface gravity ($\kappa_H$)  &  $\frac{1}{4\sigma}$  & $\frac{1}{4\sigma} e^\delta$ & $\frac{1}{4\sigma}$ & $\frac{1}{4\sigma} e^\delta$ \\[2mm]
\hline
 \rule{0pt}{3ex} Horizon area ($A_H$) & $16\pi\sigma^2$ & $16\pi\sigma^2 e^{-\delta}$ & $16\pi\sigma^2$ & $16\pi\sigma^2 e^{-\delta}$ \\[1mm]
\hline
 \rule{0pt}{3ex} Komar mass ($M_H$) & $\sigma$ & $\sigma$ & $\sigma$ & $\sigma$ \\[1mm]
\hline
 \rule{0pt}{3ex} Magnetic flux ($F_B$) & - & - &  $\frac{4\pi\sigma^2 B}{1 + B^2\sigma^2}$ & $\frac{4\pi\sigma^2 B}{1 + B^2\sigma^2}$\\[2mm]
 \hline
\end{tabular}
\end{center}
\caption{\footnotesize{Overview of the properties of the isolated Schwarzschild black hole versus its generalizations in an external gravitational field and/or an external magnetic field (see main text). The notation $\delta = 2\sum_{n=1}^\infty a_{2n}$ is used.}}\label{static}
\end{table}

\section{Magnetized Reissner-Nordstr\"{o}m black hole in an external gravitational field}

In this section we construct a magnetized Reissner-Nordstr\"{o}m black hole in an external gravitational field. It is obtained by performing a Harrison transformation on the distorted Reissner-Nordstr\"{o}m solution, which we described in section 2.2, as a seed. For the purpose we develop a construction scheme, which is valid for any seed solution belonging to the electrovacuum Weyl class, and apply it in our particular case. We represent the seed metric by means of the potential with respect to the spacelike Killing field

\begin{eqnarray}\label{metric_u}
ds^2 = e^{2u_0}d\phi^2 + \sigma^2e^{-2u_0}\left[e^{2\gamma^{'}_0} \left(\frac{dx^2}{x^2-1}+\frac{dy^2}{1-y^2}\right) - (x^2-1)(1-y^2) dt^2\right],
\end{eqnarray}
where the potential $u_0$ and the metric function $\gamma^{'}_0$ are related to the original quantities $\psi$ and $\gamma$ as

\begin{eqnarray}\label{potential_RN}
e^{2u_0}&=&\sigma^2(x^2-1)(1-y^2)e^{-2\psi}, \nonumber \\
e^{2\gamma^{'}_0} &=& \sigma^2(x^2-1)(1-y^2)e^{2\gamma-4\psi}.
\end{eqnarray}
To be consistent with the notations introduced in the description of the Harrison transformation, we rename the electromagnetic potential of the electrovacuum Weyl solution ($\ref{chi_Weyl}$) as $\chi= A^0_t$.  Then, the Ernst potential and the complex electromagnetic potential for the seed solution have the form

\begin{eqnarray}
{\cal E}_0 &=& -\sigma^2(x^2-1)(1-y^2)e^{-2\psi} -(B^0_\phi)^2, \nonumber \\
 \Phi_0 &=& iB^0_\phi,
\end{eqnarray}
where the dual electromagnetic potential $B^0_\phi$ is determined by the equations

\begin{eqnarray}\label{B_phi}
\frac{\partial B^0_\phi}{\partial x} &=& \frac{\sigma(y^2-1)}{e^{2\psi}}\frac{\partial A^0_t}{\partial y}, \nonumber \\[2mm]
\frac{\partial B^0_\phi}{\partial y} &=& \frac{\sigma(x^2-1)}{e^{2\psi}}\frac{\partial A^0_t}{\partial x}.
\end{eqnarray}

For the electrovacuum Weyl solutions the gravitational potential $\psi$ and the electromagnetic potential $A^0_t$ are functionally dependent, as shown by the relation ($\ref{chi_Weyl}$). It further leads to the differential relation

\begin{eqnarray}\label{A_psi0}
d A^0_t = -\frac{e^{2\psi}}{\sqrt{C^2-1}}\,d \psi_0,
\end{eqnarray}
where $\psi_0$ is the gravitational potential for the vacuum Weyl solution, which corresponds to the electrovacuum one by the map ($\ref{psi_Weyl}$). Then, the equations for dual potential can be easily integrated. For the distorted Reissner-Nordstr\"{o}m solution the vacuum potential $\psi_0$ is the potential of the distorted Schwarzschild solution given by ($\ref{dist_Sch}$), and the parameter $C$ is traditionally expressed by the mass and charge parameters of the Reissner-Nordstr\"{o}m solution as $C=M/Q$, while $\sigma^2 = M^2-Q^2$. Thus, we obtain for the dual potential

\begin{eqnarray}\label{B_phi0}
B^0_\phi = -Q y - Q\sum^{\infty}_{n=1}\frac{a_nnR^{n+1}}{2n+1}\left[P_{n+1}\left(\frac{xy}{R}\right) - P_{n-1}\left(\frac{xy}{R}\right)\right],
\end{eqnarray}
where we use the notations introduced in section 2. Hence, we can construct the complex function $\Lambda$ in the form

\begin{eqnarray}
\Lambda = 1-\frac{1}{4}B^2{\cal E}_0 + iBB_\phi^0.
\end{eqnarray}

In the case of a static electrovacuum seed, the equations ($\ref{omega_Mgn}$), which determine the metric function $\omega$ of the magnetized solution,   reduce to the system

\begin{eqnarray}
\frac{\partial\omega}{\partial x} &=& \frac{2\sigma(1-y^2)}{e^{2u_0}}\left[-B\,\partial_y B_\phi^0 + \frac{B^3}{4}\left(B_\phi^0\,\partial_y (e^{2u_0}) - \left[e^{2u_0} - (B_\phi^0)^2\right]\partial_y B_\phi^0\right)\right], \nonumber \\[3mm]
\frac{\partial\omega}{\partial y} &=& -\frac{2\sigma(x^2-1)}{e^{2u_0}}\left[-B\,\partial_x B_\phi^0 + \frac{B^3}{4}\left( B_\phi^0\,\partial_x (e^{2u_0}) - \left[e^{2u_0} - (B_\phi^0)^2\right]\partial_x B_\phi^0\right)\right]. \nonumber
\end{eqnarray}

\bigskip
For the electrovacuum Weyl class of solutions, they can be simplified considerably if we take into account the relations between the electromagnetic and gravitational potentials $A_t^0$, $B_\phi^0$ and $u_0$ of the seed solution, and the vacuum Weyl potential $\psi_0$ corresponding to it, determined by eqs. ($\ref{B_phi}$)-($\ref{A_psi0}$), and ($\ref{chi_Weyl}$)-($\ref{psi_Weyl}$). Thus, we obtain the following differential relation

\begin{eqnarray}
d \omega = - d\left[ 2BA^0_t +\frac{1}{2}B^3(B^0_\phi)^2\left(C-A^0_t\right) - \frac{1}{2}B^3\tilde\omega\right],
\end{eqnarray}
\noindent
where $C$ is the parameter characterizing the map between the vacuum and electrovacuum Weyl solutions. In the case of the distorted Reissner-Nordstr\"{o}m solution it is expressed by means of its mass and charge parameters as $C=M/Q$. The function $\widetilde\omega$ is a solution to the equations

\begin{eqnarray}
\frac{\partial \widetilde\omega }{\partial x} &=&-\sigma(y^2-1)^2\frac{\partial }{\partial y}\left(\frac{B^0_\phi}{1-y^2}\right), \nonumber \\[2mm]
\frac{\partial \widetilde\omega}{\partial y} &=& \sigma(x^2-1)^2\frac{\partial}{\partial x}\left(\frac{B^0_\phi}{x^2-1}\right).
\end{eqnarray}

Using the explicit form of the dual electromagnetic potential $B^0_\phi$ ($\ref{B_phi0}$) for the distorted Reissner-Nordstr\"{o}m solution we obtain

\begin{eqnarray}\label{omega1}
\widetilde\omega &=& \sigma Q\, x(1+y^2) + \sigma Q\,(x^2-1)(1-y^2)\sum^{\infty}_{n=1}\frac{a_nn(n-1)}{(n+1)(n+2)}R^{n}P_n\left(\frac{xy}{R}\right)   \nonumber\\[2mm]
&+& 2\sigma Q\,\sum^{\infty}_{n=1}\frac{a_n(n-1)}{(n+1)(n+2)}R^{n+2}\left[\frac{xy}{R}P_{n-1}\left(\frac{xy}{R}\right) - P_{n-2}\left(\frac{xy}{R}\right)\right].  \nonumber \\ \nonumber
\end{eqnarray}

Hence, the metric function $\omega$ is given by the expression

\begin{eqnarray}\label{omega}
\omega = - 2BA^0_t -\frac{1}{2}B^3(B^0_\phi)^2\left(\frac{M}{Q}-A^0_t\right) + \frac{1}{2}B^3 \widetilde\omega + const.
\end{eqnarray}

We have obtained all the quantities necessary to construct the magnetized Reissner-Nordstr\"{o}m black hole in an external gravitational field by means of the Harrison transformation. The solution is given by

\begin{eqnarray}\label{metric_M}
ds^2 &=& |\Lambda|^{-2}e^{2u_0}(d\phi + \omega dt)^2 \nonumber \\[2mm]
&+& |\Lambda|^2 \sigma^2e^{-2u_0}\left[e^{2\gamma^{'}_0} \left(\frac{dx^2}{x^2-1}+\frac{dy^2}{1-y^2}\right) - (x^2-1)(1-y^2) dt^2\right], \nonumber \\[2mm]
|\Lambda|^2 &=& (1-\frac{1}{4}B^2{\cal E}_0)^2 + B^2(B_\phi^0)^2,
\end{eqnarray}
where
\begin{eqnarray}
{\cal E}_0 &=& -e^{2u_0} - (B_\phi^0)^2, \nonumber \\[2mm]
e^{2u_0} &=& \sigma^2(x^2-1)(1-y^2)e^{-2\psi}, \quad~~~ e^{\psi} = \frac{2\sigma e^{\psi_0}}{\sigma + M + (\sigma-M)e^{2\psi_0}}, \nonumber \\[2mm]
e^{2\gamma^{'}_0} &=& \sigma^2(x^2-1)(1-y^2)e^{2\gamma-4\psi}, \nonumber \\[2mm]
A_t^0 &=& \frac{Q(1-e^{2\psi_0})}{\sigma + M + (\sigma - M)e^{2\psi_0}}, \nonumber \\[2mm]
\psi_0 &=& \frac{1}{2}\ln{\left(\frac{x-1}{x+1}\right)} + \sum^{\infty}_{n=0}a_nR^n P_n\left(\frac{xy}{R}\right), \nonumber\\[2mm]
\gamma &=&  \frac{1}{2}\ln(x^2-1)+ \sum^{\infty}_{n,k=1}\frac{nk}{n+k}a_na_k R^{n+k}\left(P_n P_k - P_{n-1}P_{k-1}\right) \nonumber \\
&+& \sum^{\infty}_{n=1}a_n\sum^{n-1}_{k=0}\left[(-1)^{n-k+1}(x+y) -x+y\right]R^k P_k\left(\frac{xy}{R}\right). \nonumber
\end{eqnarray}
The potentials $B_\phi^0$ and $\omega$ are determined by ($\ref{B_phi0}$) and ($\ref{omega}$), and $B$ is the parameter characterizing the external magnetic field. The electromagnetic field possesses the form

\begin{eqnarray}
F &=& dA_t\wedge dt + dA_\phi\wedge d\phi,  \nonumber \\
A_\phi &=& \frac{B}{2|\Lambda|^2} \left[-{\cal E}_0 + \frac{1}{4}B^2{\cal E}^2_0 + 2(B^0_\phi)^2\right],  \nonumber \\
A_t &=& 2A^0_t -\omega A_\phi + \frac{3}{2B}\omega,
\end{eqnarray}
in the same notations as in the expression for the metric, leading to the co-rotating potential $\Sigma = A_t + \omega A_\phi$

\begin{eqnarray}\label{Sigma}
\Sigma = \frac{3}{2B}\omega + 2A^0_t + const. \, ,
\end{eqnarray}
according to the general expression ($\ref{Sigma_gen}$).

\subsection{Properties}

The physical properties of the constructed solution depend on the behaviour of the potentials ${\cal E}_0$, $\widetilde\omega$,  $A_t^0$ and $B_\phi^0$, which are included in the metric functions, on the horizon $x=1$ and on the symmetry axis $y = \pm 1$. They reduce to the following expressions

\begin{eqnarray}
&&B^0_\phi(x=1) = -Qy, \quad~~~A^0_t(x=1) = \frac{Q}{\sigma + M}, \quad~~~ \widetilde\omega(x=1)= \sigma Q (1+y^2), \nonumber \\[2mm]
&&{\cal E}_0 (x=1)= (1-y^2)(\sigma + M)^2\exp\left(\sum^{\infty}_{n=1}a_n y^n\right)-Q^2y^2, \nonumber \\[2mm]
&&B^0_\phi(y=\pm 1) = \mp Q,  \quad~~~ {\cal E}_0 (y=\pm 1)= -Q^2.
\end{eqnarray}

\noindent
We see that most of the quantities are not influenced by the external gravitational field, and coincide with those for the magnetized Reissner-Nordstr\"{o}m black hole. This will result in similarities in the physical behavior of the solutions.

The constructed magnetized solution is stationary rather than static. The Harrison transformation introduces rotation, which is interpreted as caused by the interaction of the intrinsic electromagnetic potential of the seed solution with the external magnetic field. The horizon rotates with angular velocity $\Omega_H = -\omega(x=1)$ with respect to the symmetry axis given by

\begin{eqnarray}\label{Omega}
\Omega_H = \frac{2BQ}{\sigma+M}-\frac{1}{2}B^3Q\sigma + C_\omega.
\end{eqnarray}

The angular velocity is determined up to an arbitrary constant $C_\omega$. For asymptotically flat solutions this freedom is fixed by the requirement to avoid global rotation of the spacetime. However, since we consider a local solution, there is no natural constraint to be imposed without considering an extension to an asymptotically flat solution. For some asymptotically non-flat solutions the constant can be fixed by the requirement that the metric function  $\omega$ vanishes on the rotational axis, as for example for rotating black holes in an external gravitational field \cite{Breton:1997}, \cite{Nedkova:2014}. However, such normalization cannot be applied for magnetized solutions, since the metric function $\omega$ does not reduce to a constant on the axis.

The horizon angular velocity ($\ref{Omega}$) is not influenced by the external gravitational field.  It coincides with the angular velocity of the magnetized Reissner-Nordstr\"{o}m black hole, and the expression given in \cite{Gibbons:2013} is recovered by setting the constant $C_\omega = -\frac{1}{2}B^3QM$\footnote{In the case of the magnetized Reissner-Nordstr\"{o}m black hole the global rotation of the spacetime cannot be avoided, and the value of the angular velocity of the horizon is determined up to a constant.}.

\begin{eqnarray}
\Omega_H = \frac{2BQ}{\sigma+M}\left[1-\frac{1}{4}B^2(\sigma+M)^2\right].
\end{eqnarray}
In order to obtain a solution, which is free of conical singularities, we should examine its behavior near the rotational axis. We can require that the standard periodicity of the angular coordinate $\Delta\phi = 2\pi$ is preserved. Then,  conical singularities are avoided provided that the norm $K = g(\eta,\eta)$ of the spacelike Killing vector $\eta=\partial/\partial\phi$ satisfies the condition

\begin{eqnarray}\label{conicalG}
\frac{1}{4K}g^{\mu\nu}\partial_\mu K\partial_\nu K \longrightarrow 1,
\end{eqnarray}
in the vicinity of the rotational axis. Explicit calculation shows that it reduces to the requirement

\begin{eqnarray}\label{conical_MRN}
e^{\widetilde\gamma_0} |\Lambda|^2\mid_{y=\pm 1} = \exp\left(\sum^{\infty}_{n=0}a_{2n+1}\right)\left[\left(1 + \frac{1}{4}B^2Q^2\right)^2 + B^2Q^2\right] =1.
\end{eqnarray}
\noindent
This relation introduces a dependence between the parameters $a_n$ characterizing the external gravitational field, and the external magnetic field parameter $B$, and it can be satisfied only if some of the parameters $a_n$ for odd $n$ possess negative values. Still, regular solutions with standard periodicity $\Delta\phi= 2\pi$ can be achieved for special cases of the external gravitational potential. Another possibility to obtain solutions which are free of conical singularities is to assign to the angular coordinate the periodicity

\begin{eqnarray}\label{Delta_phi}
\Delta\phi = 2\pi \exp\left(\sum^{\infty}_{n=0}a_{2n+1}\right)\left[\left(1 + \frac{1}{4}B^2Q^2\right)^2 + B^2Q^2\right].
\end{eqnarray}
In this case the parameters characterizing the external gravitational and magnetic fields remain independent. For the magnetized Reissner-Nordstr\"{o}m black hole conical singularities can be avoided only by choosing an appropriate periodicity of the angular coordinate, which we denote by $\Delta\phi_M$

\begin{eqnarray}\label{Delta_phiM}
\Delta\phi_M = 2\pi \left[\left(1 + \frac{1}{4}B^2Q^2\right)^2 + B^2Q^2\right].
\end{eqnarray}

We calculate further the surface gravity of the horizon defined as
\begin{eqnarray}
\kappa^2_H = -\frac{1}{4\lambda} g^{\mu\nu}\partial_{\mu}\lambda
\partial_{\nu}\lambda ,
\end{eqnarray}

\noindent
where $\lambda=g(V,V)$ is the norm of the Killing field $V = \partial/\partial t + \Omega_H\partial/\partial\psi$, which becomes null on the horizon.  We obtain the expression

\begin{eqnarray}
\kappa_H =\frac{\sigma}{(\sigma+ M)^2} \exp\left[2\sum_{n=1}^\infty a_{2n}\right],
\end{eqnarray}
which coincides with that of the distorted Reissner-Nordstr\"{o}m solution ($\ref{dist_RN}$). The restriction of the metric on the horizon cross-section is given by
\begin{eqnarray}
ds_H^2 &=&  (\sigma+ M)^2e^{\widetilde\gamma_0(y)-2\widetilde\psi_0(y)}\left[|\Lambda|^{-2}e^{-\widetilde\gamma_0(y)}(1-y^2)d\phi^2 + |\Lambda|^2 e^{\widetilde\gamma_0(y)}\frac{dy^2}{1-y^2}\right]  \\[2mm]
 &=& (\sigma+ M)^2 e^{\widetilde\gamma_0(\theta)-2\widetilde\psi_0(\theta)}\left[|\Lambda|^{-2}e^{-\widetilde\gamma_0(\theta)}\sin^2\theta d\phi^2 + |\Lambda|^2e^{\widetilde\gamma_0(\theta)}d\theta^2\right], \quad~~~ y = \cos\theta. \nonumber
\end{eqnarray}
The horizon radius is not influenced by the magnetic field, and coincides with that of the vacuum seed solution ($\ref{dist_RN}$)

\begin{eqnarray}
R_H &=& (\sigma+M)\exp{\left[-\sum^{\infty}_{n=0}a_{2n}\right]}.
\end{eqnarray}
The deformation of the horizon due to the magnetic field is reflected in the appearance of the factor $|\Lambda|^2$ in the shape function. The horizon area given by $A_H = 2\Delta\phi R_H^2$ coincides with the distorted Reissner-Nordstr\"{o}m solution provided the regularity condition ($\ref{conical_MRN}$) is satisfied, and deviates by the factor ($\ref{Delta_phi}$) otherwise.

The constructed magnetized black hole possesses the electric charge

\begin{eqnarray}\label{charge}
{\cal Q}_H = \frac{1}{4\pi}\int_{H} \star F = \frac{\Delta\phi}{\Delta\phi_M}Q\left(1-\frac{1}{4}B^2Q^2\right),
\end{eqnarray}
while the restriction of the co-rotating potential $\Sigma$ on the horizon is given by
\begin{eqnarray}
\Sigma_H = -\frac{Q}{\sigma+M}+\frac{3}{4}B^3Q\sigma + C_\Sigma.
\end{eqnarray}
The co-rotating potential $\Sigma_H$ suffers from the same ambiguity as the angular velocity, which is connected with the choice of the arbitrary constant $C_\Sigma$. The expression coincides with the corresponding quantity for the magnetized Reissner-Nordstr\"{o}m solution, reflecting no influence of the external gravitational field. Setting $C_\Sigma = -\frac{3}{4}B^3Q M$ leads to the value proposed in \cite{Gibbons:2013b}

\begin{eqnarray}
\Sigma_H = -\frac{Q}{\sigma+M}\left[1 - \frac{3}{4}B^2(\sigma + M)^2\right].
\end{eqnarray}

The solution is further characterized by the magnetic flux through the horizon upper/lower hemisphere
\begin{eqnarray}\label{M_flux}
F_B &=& \frac{1}{4\pi} \int_{H^+} F = \frac{\Delta\phi}{4\pi} \left[ A_\phi(x=1, y=0) - A_\phi(x=1, y=1)\right]  \nonumber \\[2mm]
 &=& \frac{\Delta\phi}{\Delta\phi_M} \frac{B(\sigma + M)}{2|\Lambda|^2_{y=\pm 1}}\frac{\left[(2\sigma-M) + \frac{1}{4}B^2q^2(2\sigma+M)\right]}{1 + \frac{1}{4}B^2(\sigma+M)^2}, \nonumber \\[3mm]
 &&|\Lambda|^2_{y=\pm 1} = \left(1 + \frac{1}{4}B^2Q^2\right)^2 + B^2Q^2.
\end{eqnarray}

The expressions for the charge and the magnetic flux deviate from the corresponding quantities  for the magnetized Reissner-Nordstr\"{o}m black hole (see e.g. \cite{Aliev:1989}) only by the modification of the angular coordinate periodicity  $\Delta\phi$ with respect to that of the magnetized solution $\Delta\phi_M$ given by ($\ref{Delta_phiM}$). Similar to the magnetized Reissner-Nordstr\"{o}m black hole the solution does not exhibit the Meissner effect in the extremal horizon limit for general values of the parameter $B$. Besides the component sourced by the external magnetic field, the magnetic flux through the horizon contains an intrinsic part generated by the rotation of the charged black hole. The second contribution does not vanish in the extremal limit, and therefore flux expulsion is not observed for general extremal horizons. The only possibility to obtain vanishing magnetic flux is to tune the external field parameter $B$ in such a way that the physical charge of the black hole ($\ref{charge}$) is set to zero, i.e. $ B = \pm 2/Q$ \cite{Gibbons:2013}. This condition does not prevent extremal solutions, which are obtained in the limit $\sigma = M$. Inserting the zero-charge condition in eq. ($\ref{M_flux}$), as well as the extremality condition, we see that the magnetic flux threading the horizon vanishes.

We summarize the examined properties in table $\ref{RN}$, where we compare their behavior for the isolated Reissner-Nordstr\"{o}m black hole (Isol. R.N.), its generalizations in an external gravitational field \cite{Fairhurst} (Dist. R.N.), and in an external magnetic field \cite{Ernst:1976a} (Magn. R.N.), and for the constructed solution in this section (Magn. Dist. R.N.), which reflects the influence of both external potentials.

\begin{table}[h] \small{
\begin{center}
\begin{tabular} {|c|c|c|c|c|}
\hline
 &\rule{0pt}{2ex} Isol. R.N. &  Dist. R.N.& Magn. R.N.&  Magn. Dist. R.N.\\[1mm]
\hline
\rule{0pt}{3ex}   Surface gravity &  $\frac{\sigma}{r^2_{+}}$  & $\frac{\sigma}{r^2_{+}} e^\delta$ & $\frac{\sigma}{r^2_{+}}$ & $\frac{\sigma}{r^2_{+}} e^\delta$ \\
($\kappa_H$)&&&& \\[1mm]
\hline
 \rule{0pt}{3ex} Horizon area & $4\pi r^2_{+}$ & $4\pi r^2_{+} e^{-\delta}$ & $2\Delta\phi_M r^2_{+}$ & $2\Delta\phi r^2_{+} e^{-\delta}$ \\
 ($A_H$)&&&& \\[1mm]
\hline
 \rule{0pt}{3ex}\footnotesize{ Angular velocity } & - & - & $\frac{2BQ}{r_{+}}\left[1-\frac{1}{4}B^2r_{+}^2\right]$ & $\frac{2BQ}{r_{+}}\left[1-\frac{1}{4}B^2r_{+}^2\right]$ \\
 ($\Omega_H$)&&&& \\[1mm]
\hline
\rule{0pt}{3ex} EM potential  & $-\frac{Q}{r_{+}}$ & $-\frac{Q}{r_{+}}$& $-\frac{Q}{r_{+}}\left[1-\frac{3}{4}B^2r_{+}^2\right]$ & $-\frac{Q}{r_{+}}\left[1-\frac{3}{4}B^2r_{+}^2\right]$ \\
($\Sigma_H$)&&&& \\[1mm]
\hline
\rule{0pt}{3ex} Electric charge & $Q$ & $Q$ & $Q\left[1-\frac{1}{4}B^2Q^2\right]$ & $\frac{\Delta\phi}{\Delta\phi_M}Q\left[1-\frac{1}{4}B^2Q^2\right]$ \\
(${\cal Q}_H$)&&&& \\[1mm]
\hline
 \rule{0pt}{3ex} Magnetic flux  & - & - &  $\frac{\Delta\phi_M}{\Delta\phi} F_B$ & $F_B$\\
($F_B$)&&&& \\[1mm]
 \hline
\end{tabular}
\end{center}
\caption{\footnotesize{Overview of the properties of the isolated Reissner-Nordstr\"{o}m black hole versus its generalizations in an external gravitational field and/or an external magnetic field. The notations $\delta = 2\sum_{n=1}^\infty a_{2n}$ and $r_+ = \sigma + M$ are used. In the normalization of the angular velocity and the electrostatic potential we adopt the same convention as in \cite{Gibbons:2013},\cite{Gibbons:2013b} (see main text). The magnetic flux $F_B$ refers to the quantity obtained in eq. ($\ref{M_flux}$). }}\label{RN}}
\end{table}

\subsubsection{Local mass and angular momentum}

We can calculate the local mass and angular momentum of the black hole by using Wald's procedure for calculating conserved charges. In Einstein-Maxwell gravity every Killing field $k$ generates a Noether current ${\cal J}$ given by \cite{Wald:1993a}, \cite{Wald:1993b}.

\begin{eqnarray}
{\cal J} = - d\star dk - 4\star F\wedge d(k^\mu A_\mu),
\end{eqnarray}
where $F$ is the Maxwell 2-form, and $A_\mu$ is the electromagnetic potential. It satisfies $d{\cal J} =0$ for solutions of the field equations, and therefore we can express it as ${\cal J} = -d{\cal P}$ by means of a Noether 2-form ${\cal P}$. Integrating the 2-form ${\cal P}$ over a closed surface $\Sigma$ we define the Noether charge of $\Sigma$ relative to the Killing filed $k$

\begin{eqnarray}
{\cal Q}[k] =  \int_\Sigma {\cal P}.
\end{eqnarray}

For stationary and axisymmetric solutions of the Einstein-Maxwell equations we can construct the Noether charges associated with the Killing fields generating the azimuthal symmetry and the time translations, or in our notations $\eta = \partial/\partial\phi$ and $\chi=\partial/\partial t$. Integrating the corresponding Noether 2-forms ${\cal P}_\eta$ and ${\cal P}_\chi$ over the horizon of our solution ($\ref{metric_M}$), we obtain the local angular momentum ${\cal J}_H$ and the local mass ${\cal M}_H$ of the black hole

\begin{eqnarray}
{\cal J}_H &=& -\frac{1}{16\pi} \int_H {\cal P}_\eta = \frac{1}{16\pi} \int_H \star d\eta + 4 A_\phi \star F, \nonumber \\[2mm]
{\cal M}_H &=& \frac{1}{8\pi} \int_H {\cal P}_\chi = -\frac{1}{8\pi} \int_H \star d\chi + 4 A_t \star F.
\end{eqnarray}

By construction, the two quantities are connected by the Smarr relation

\begin{eqnarray}\label{Smarr}
{\cal M}_H = \frac{1}{4\pi}\kappa_H A_H + 2\Omega_H{\cal J}_H + \Sigma_H{\cal Q}_H,
\end{eqnarray}
where all the thermodynamic quantities were previously defined. Hence, obtaining the angular momentum determines also the horizon mass through ($\ref{Smarr}$).

The calculation of the angular momentum can be accomplished by performing a dimensional reduction on the $\phi$ coordinate as proposed in
\cite{Gibbons:2013b}. Then, the integral reduces to
\begin{eqnarray}
{\cal J}_H = -\frac{1}{16\pi} \int_H {\cal P}_\eta = \frac{(\Delta\phi)^2}{16\pi^2}\int_H d \lambda = \frac{(\Delta\phi)^2}{16\pi^2}\lambda|^{y=1}_{y=-1}, \end{eqnarray}
where $\lambda$ is a potential defined in the factor space. The explicit form of the potential $\lambda$ for solutions generated by means of the Harrison transformation is given in \cite{Gibbons:2013}.\footnote {In \cite{Gibbons:2013} the potential is denoted by $\sigma$.} In the case when the transformation is applied to a static seed with Ernst potential ${\cal E}_0$ and electromagnetic potential $\Phi_0 = i B_\phi^0$, it reduces to the expression
\begin{eqnarray}
\lambda = -\frac{B B_\phi^0}{|\Lambda|^4}\left[({\cal E}_0 - (B_\phi^0)^2)(1 -\frac{1}{4}B^2{\cal E}_0^2)\right] + 2A_\phi B_\phi,
\end{eqnarray}
in the notations we previously introduced. The potential $\lambda$ is not invariant with respect to gauge transformations of the electromagnetic potential. Therefore, it was suggested in \cite{Gibbons:2013b} to define the angular momentum by means of a related potential $\tilde\lambda = \lambda - 2 A_\phi B_\phi$. Still, both definitions lead to equivalent results if the electromagnetic potential $A_\phi$ is normalized so that it vanishes on the symmetry axis. In this normalization we obtain for the horizon angular momentum

\begin{eqnarray}
{\cal J}_H = -\frac{(\Delta\phi)^2}{(\Delta\phi_M)^2}\, Q^3 B(1 +\frac{1}{4}B^2Q^2),
\end{eqnarray}
where $\Delta\phi_M$ is the periodicity of the angular coordinate for the magnetized Reissner-Nordstr\"{o}m solution. The horizon angular momentum deviates from the corresponding quantity for the magnetized Reissner-Nordstr\"{o}m solution calculated by the same procedure, only by the difference in the periodicities $\Delta\phi$ and $\Delta\phi_M$. The local mass ${\cal M}_H$ is not determined uniquely. Since the horizon angular velocity $\Omega_H$ and the co-rotating potential $\Sigma_H$ are determined up to an arbitrary additive constant, the same ambiguity is inherited in the value of the horizon mass.

\section{Conclusion}

Black holes in  astrophysical environments are involved in gravitational and electromagnetic interaction with surrounding matter fields. Usually they are considered negligible compared to the central potential, and the spacetime is described by an isolated black hole solution. Yet, certain gravitational phenomena are sensible to even small perturbations, and lead to qualitatively different observable features. Valuable intuition in this respect can be provided by exact solutions which include the back-reaction of the black hole geometry to an external gravi-magnetic potential.

In this paper we constructed a family of stationary black hole solutions  reflecting the interaction with an external matter distribution, which produces a strong  magnetic field. The solutions are obtained by a Harrison transformation, and could model to some extent the conditions in the vicinity of a black hole surrounded by an accretion disk. We investigated their thermodynamical properties making comparison with the corresponding isolated black holes, as well as with the cases when only the gravitational or electromagnetic interaction is taken into account. We also studied the influence of the external gravitational potential on the Meissner effect, and demonstrated that the same qualitative behavior is observed as for purely magnetized solutions.

\section*{Acknowledgment}
J.K. and P.N. gratefully acknowledge support by the DFG Research Training Group 1620 ``Models of Gravity''.  P.N. and S.Y. are partially supported by the Bulgarian NSF Grant $\textsl{DFNI T02/6}$.

\end{document}